\documentclass[prl,aps,twocolumn,amsmath,amssymb]{revtex4}

\def\gsim{ \lower .75ex \hbox{$\sim$} \llap{\raise .27ex \hbox{$>$}} }

\def\lsim{ \lower .75ex \hbox{$\sim$} \llap{\raise .27ex \hbox{$<$}} }

\def\ie{{\it i.e.\ }}
\def\IZ{\relax\ifmmode\mathchoice
{\hbox{\cmss Z\kern-.4em Z}}{\hbox{\cmss Z\kern-.4em Z}}
{\lower.9pt\hbox{\cmsss Z\kern-.4em Z}} {\lower1.2pt\hbox{\cmsss
Z\kern-.4em Z}}\else{\cmss Z\kern-.4em Z}\fi}
\def\IR{\relax{\rm I\kern-.18em R}}

\def\one{{\hbox{ 1\kern-.8mm l}}}

\newlength{\bredde}
\def\slash#1{\settowidth{\bredde}{$#1$}\ifmmode\,\raisebox{.15ex}{/}
\hspace*{-\bredde} #1\else$\,\raisebox{.15ex}{/}\hspace*{-\bredde}
#1$\fi}
\newcommand{\ft}[2]{{\textstyle\frac{#1}{#2}}}

\newsavebox{\zzzbar}
\sbox{\zzzbar}
  {\setlength{\unitlength}{0.9em}
  \begin{picture}(0.6,0.7)
  \thinlines
  \put(0,0){\line(1,0){0.6}}
  \put(0,0.75){\line(1,0){0.575}}
  \multiput(0,0)(0.0125,0.025){30}{\rule{0.3pt}{0.3pt}}
  \multiput(0.2,0)(0.0125,0.025){30}{\rule{0.3pt}{0.3pt}}
  \put(0,0.75){\line(0,-1){0.15}}
  \put(0.015,0.75){\line(0,-1){0.1}}
  \put(0.03,0.75){\line(0,-1){0.075}}
  \put(0.045,0.75){\line(0,-1){0.05}}
  \put(0.05,0.75){\line(0,-1){0.025}}
  \put(0.6,0){\line(0,1){0.15}}
  \put(0.585,0){\line(0,1){0.1}}
  \put(0.57,0){\line(0,1){0.075}}
  \put(0.555,0){\line(0,1){0.05}}
  \put(0.55,0){\line(0,1){0.025}}
  \end{picture}}

\newcommand{\ena}{\end{eqnarray}}
\newcommand{\beqa}{\begin{eqnarray}}
\newcommand{\eeqa}{\end{eqnarray}}
\newcommand{\bea}{\begin{eqnarray}}
\newcommand{\eea}{\end{eqnarray}}

\newcommand{\be}{\begin{equation}}
\newcommand{\ee}{\end{equation}}
\def\ft#1#2{{\textstyle{\frac{\scriptstyle #1}{\scriptstyle #2} } }}
\def\fft#1#2{{\frac{#1}{#2}}}

\usepackage{graphicx}

\def\nn{\nonumber}
\def\ben{\begin{equation}}
\def\een{\end{equation}}

\def\bea{\begin{eqnarray}}
\def\eea{\end{eqnarray}}
\def\be{\begin{equation}}
\def\ee{\end{equation}}
\def\beq{\begin{eqnarray}}
\def\eeq{\end{eqnarray}}

\def\({\left (}
\def\){\right )}
\def\[{\left [}
\def\[{\right ]}

\def\ba{\begin{eqnarray}}
\def\ea{\end{eqnarray}}

\def\del{{\partial}}

\input amssym.def
\input amssym.tex

\usepackage{dcolumn}

\begin{document}

\begin{flushright}
CAQS-1501 \ \  Imperial/TP/15/KSS/01 \ \  MI-TH-1504
\end{flushright}

\title{Black Holes in Higher-Derivative Gravity }

\author{H. L\"u$^1$,  A. Perkins$^2$, C.N. Pope$^{3,4}$ and K.S. Stelle$^2$}

\affiliation{\vskip 0.3cm
{\it $^1$ Center for Advanced Quantum Studies, 
Department of Physics, Beijing Normal University,
Beijing 100875, China}
\vskip .5mm
{\it $^2$ The Blackett Laboratory, Imperial College London,
Prince Consort Road, London SW7 2AZ, UK}
\vskip 0.5mm
{\it $^3$ George P. \& Cynthia W. Mitchell Institute for Fundamental Physics and Astronomy, Texas A\&M University, College Station, TX 77843-4242, USA}
\vskip .5mm
{\it $^4$ DAMTP, Centre for Mathematical Sciences, Cambridge University, Wilberforce Road, Cambridge, CB3 0WA, UK }
\vskip 0.5mm}

\begin{abstract}

  Extensions of Einstein gravity with higher-order derivative
terms arise in string theory and other effective theories, as well as being
of interest in their own right.  In this paper we
study static black-hole solutions in the example of Einstein gravity with
additional quadratic curvature terms.  A Lichnerowicz-type theorem
simplifies the analysis by establishing that they must have vanishing
Ricci scalar curvature.  By numerical methods we then demonstrate the
existence of further black-hole solutions over and above the Schwarzschild
solution.  We discuss some of their thermodynamic properties, and show
that they obey the first law of thermodynamics.

\end{abstract}

%\pacs{PACS number(s): 98.80.Es, 98.80.Cq, 03.70.+k}

\maketitle

%\begin{multicols}{2}[]

   The well-known problem of the non-renormalisability of Einstein gravity
has given rise to many attempts to view it as an effective low-energy theory
that will receive higher-order corrections that become important as the
energy scale increases (see, for example, \cite{thoovelt}).  
In string theory, the Einstein-Hilbert action
is just the first term in an infinite series of gravitational
corrections built from powers of the curvature tensor and its
derivatives.  In other approaches, only a
finite number of additional terms might be added.  
It was shown in \cite{stelle1} that if one adds all possible quadratic
curvature invariants to the usual Einstein-Hilbert action one obtains a
renormalisable theory, albeit at the price of introducing ghost-like
modes.  Arguments have been given for why these might not be 
fatal to the theory (for example, see \cite{smilga} for a recent 
discussion).  
In any case, it is worthwhile to study in detail
the properties of the theory of Einstein gravity with 
added quadratic curvature terms, in order shed light on the 
question of whether it has irredeemable
pathologies or whether they can be controlled in some manner.

   Black holes are the most fundamental objects in a
theory of gravity, and they provide powerful probes for studying some of the
more subtle global aspects of the theory.  It is therefore of
considerable interest to investigate the structure of black-hole
solutions in theories of gravity with higher-order curvature terms. In this
paper, we report on some investigations of the static,
spherically-symmetric black-hole solutions in four-dimensional 
Einstein-Hilbert gravity with
added quadratic curvature terms, for which the most general action can be
taken to be
%%%%%
\be
\label{HDGaction}
I = \int d^4x\sqrt{-g}\left(\gamma R -
\alpha C_{\mu\nu\rho\sigma}C^{\mu\nu\rho\sigma} + \beta R^{2}\right)\,,
\ee
%%%%%
where $\alpha$, $\beta$ and $\gamma$ are constants and
$C_{\mu\nu\rho\sigma}$ is the Weyl tensor.  We shall work in units where
we set $\gamma=1$, and the equations of motion following from
(\ref{HDGaction}) are then
%%%%%
\bea
&&R_{\mu\nu}-\ft12 R g_{\mu\nu} - 4 \alpha B_{\mu\nu} +2\beta
 R(R_{\mu\nu}-\ft14 R g_{\mu\nu}) \nn\\
&& + 2\beta(g_{\mu\nu}\square R-\nabla_\mu\nabla_\nu R)=0\,,\label{eoms}
\eea
%%%%%
where $B_{\mu\nu}= (\nabla^\rho\nabla^\sigma +
    \ft12 R^{\rho\sigma}) C_{\mu\rho\nu\sigma}$ is the Bach tensor,
which is tracefree.

In general, the theory describes a system with
a massive spin-2 mode with mass-squared $m_2^2=1/(2\alpha)$ and
 a massive spin-0 mode with mass-squared $m_0^2=1/(6\beta)$, in addition
to the massless spin-2 graviton.  These massive modes will be associated
with rising and falling Yukawa type behaviour in the metric modes near
infinity \cite{stelle2}, of the form $\ft1{r} e^{\pm m_2 r}$ and
 $\ft1{r} e^{\pm m_0 r}$.  In particular, one can expect that if generic
initial data is set at some small distance, the rising exponentials will
eventually dominate, leading to singular asymptotic behaviour.  In seeking
black-hole solutions, the question then arises as to whether the rising
exponentials can be avoided for appropriately finely-tuned initial data.

 It can easily be seen that any solution of
pure Einstein gravity will also be a solution of
(\ref{eoms}), and so in particular the usual Schwarzschild black hole
continues to be a solution in the higher-order theory.  The question we
wish to address, then, is whether there exist any other static black
hole solutions, over and above the Schwarzschild solution.

   Static, spherically-symmetric black-hole solutions have been investigated
in \cite{nelson}, using generalisations of the Lichnerowicz and
Israel theorems for Einstein gravity.  Since we will arrive at somewhat
different conclusions, we shall briefly summarise the
key elements in \cite{nelson}, although derived in a different
notation.  We consider static metrics of the form
$ds_4^2 = -\lambda^2\, dt^2 + h_{ij}\,dx^i dx^j$,
where $\lambda$ and $h_{ij}$ are functions only of the three spatial
coordinates $x^i$. 
Taking the trace of the field equations (\ref{eoms}) gives
$\beta\, (\square- m_0^2) R=0$.
We then multiply this by
$\lambda R$ and integrate over the spatial domain from a putative
horizon out to infinity.  Expressed in terms of the covariant derivative
$D_i$ with respect to the spatial 3-metric $h_{ij}$, this gives
%%%%%
\be
\int \sqrt{h}\, d^3 x\Big[D^i(\lambda R D_i R) - \lambda (D_i R)^2 -
    m_0^2 \lambda R^2\Big]=0\,.
\ee
%%%%%
Since $\lambda$ vanishes on the horizon, it follows that if
$D_i R$ goes to zero sufficiently rapidly at spatial infinity the total
derivative (\ie surface term) gives no contribution, and the non-positivity
of the remaining terms then implies $R=0$.  In other words, as
shown in \cite{nelson}, any static black-hole solution of
(\ref{HDGaction}) must have vanishing Ricci scalar.
This leads to a great simplification, and it
means that one can, without loss of generality, study the case
of pure Einstein-Weyl gravity (\ie (\ref{HDGaction}) with $\beta=0$),
since obviously the term quadratic in $R$ makes no contribution to the field
equations for a configuration with $R=0$.  Furthermore, the trace
of the field equations (\ref{eoms})
for Einstein-Weyl gravity immediately implies $R=0$.  In fact, the two
differential equations for $h$ and $f$ are both now of only second order
in derivatives.

The second stage of the discussion in \cite{nelson} 
then involved looking at the
remaining content of (\ref{eoms}), \ie the non-trace part.
According to \cite{nelson}, this led to another integral identity that
then implied, under certain assumptions, that $R_{\mu\nu}=0$.
If this were correct, then the conclusion
would be that the usual Schwarzschild solution is the only static black
hole solution of the theory described by (\ref{HDGaction}).  However,
we find that there are sign errors in the expression given in \cite{nelson}.
Setting $R=0$, as already argued above, multiplying (\ref{eoms}) by
$\lambda R^{\mu\nu}$, and then integrating over the
spatial region outside the horizon gives
%%%%%
\bea
&&\int\sqrt{h} d^3 x\Big[D^i W_i -
 \ft14 \lambda (D_i \bar R - 4 D^j R_{ij})^2
+ 4\lambda (D^j R_{ij})^2\nn\\
&& -4\lambda (D_{[i} R_{j]k})^2 +
  \lambda (D_i R_{jk})^2
- \ft14 \lambda \bar R^2 (m_2^2 + \bar R)\nn\\
&&
  -\lambda(m_2^2 R^{ij} R_{ij} - 2 R^{ij} R_{jk} R^k{}_i)\Big]=0\,,
\eea
%%%%%
where $W_i= \lambda R^{jk} D_i R_{jk} + \ft14\lambda \bar R D_i \bar R
  -2\lambda R^{jk} D_j R_{ik} - \lambda \bar R D_j R_i{}^j$, and
$\bar R$ is the Ricci scalar of the spatial metric $h_{ij}$.
Although
the surface term will give zero, the mix of positive and negative signs in the
bulk terms precludes one from obtaining any kind of vanishing theorem for
the Ricci tensor of the four-dimensional metric.  This raises the
intriguing possibility that there might in fact exist static,
spherically symmetric black-hole solutions over and above the Schwarzschild
solution.

The equations of motion following from (\ref{HDGaction}) are too
complicated to be able to solve explicitly, even for the
case of the static, spherically-symmetric ansatz
%%%%%
\be
ds^2 = -h(r) dt^2 + \fft{dr^2}{f(r)} + r^2 (d\theta^2 + \sin^2\theta
d\phi^2)\,.\label{metans}
\ee
%%%%%
In our work, we have therefore carried out a numerical investigation of the
solutions.   To do this, we begin by supposing that there exists
a black-hole horizon at some radius $r=r_0>0$, at which the metric
functions $h$ and $f$ vanish, and we then obtain  near-horizon
Taylor expansions for $h(r)$ and $f(r)$, of the form
%%%%%
\bea
h(r) &=& c\, \Big[ (r-r_0) + h_2\, (r-r_0)^2 + h_3\, (r-r_0)^3
                  +\cdots\Big]\,,\nn\\
f(r) &=& f_1\, (r-r_0) + f_2\, (r-r_0)^2 + f_3\, (r-r_0)^3 +\cdots
\label{hfexp}
\eea
%%%%%
Substituting into the equations of motion (\ref{eoms}), with $\beta$
set to zero for the reasons discussed above, the coefficients $h_i$ and
$f_i$ for $i\ge 2$ can be solved for in terms of the two non-trivial
free parameters $r_0$ and $f_1$.  There is also
a ``trivial'' parameter, corresponding to the freedom to rescale the
time coordinate, which we have accordingly written in the form of an
overall scaling of $h(r)$.   Thus we have
%%%%%
\be
h_2=\fft{1 - 2f_1\, r_0}{f_1\, r_0^2} +
   \fft{1-f_1\, r_0}{8\alpha f_1^2\, r_0}\,,\ \ \
f_2= \fft{1-2 f_1\, r_0}{r_0^2} -\fft{3(1-f_1\, r_0)}{8 \alpha f_1\, r_0}\,,\nn
\ee
%%%%%
and so on.  (We used Taylor expansions to ${\cal O}((r-r_0)^9)$
in our numerical integrations.)
The Schwarzschild solution corresponds to $f_1=1/r_0$, and
so it is convenient to parameterise $f_1$ as
%%%%%
\be
f_1 = \fft{1+\delta}{r_0}\,,
\ee
%%%%%
with non-vanishing $\delta$ characterising the extent to which the
near-horizon solution deviates from Schwarzschild.

We
use the expansions to set initial data at a radius $r_i$ just outside
the horizon, and then use numerical routines in Mathematica to integrate the
equations out to large radius.  Generically, one finds that for a
given choice of the parameters $r_0$ and $\delta$ the solution rapidly
becomes singular as one integrates outwards from $r=r_i$, as expected
in view of our earlier observations about the rising Yukawa terms in
the asymptotic form for the metric.  If we fix a
particular value for $r_0$, we can then use the
shooting method to try to home in on a special value of $\delta$ for which
the outward integration can proceed without encountering a singularity. Of
course in practice, because of accuracy limitations in the integrations,
the solution will always eventually become singular at large enough $r$.
The signal for a good black-hole solution is that 
$f(r)$ and $h(r)$ should approach constants as $r$ increases, 
(in fact Ricci scalar flatness implies $f(r)$ must approach 1), 
and that by stepping up the accuracy
and precision goals in the calculations one can extend at will
the maximum upper limit $r=r_f$ for which the smooth behaviour can be
achieved. In practice, by running the routines with accuracy and
precision goals of order 20 decimal places, we have been able to
obtain very clean and trustworthy solutions out to at least 60 times the
horizon radius.

   Our findings are that there exists a range of values for the horizon
radius, bounded below by a certain multiple of the length $\sqrt{\alpha}$,
for which we can obtain precisely one static black-hole
solution in addition to the Schwarzschild solution.  In order to make
the statement of our results in the most concise possible way, it is
convenient, without loss of mathematical generality, to make a specific
choice for the value of $\alpha$ in (\ref{HDGaction}).  We shall take
%%%%%
\be
\alpha=\fft12\,.
\ee
%%%%%
We then find that for each choice of $r_0> r_0^{\rm min}$, where
%%%%%
\be
r_0^{\rm min} \approx 0.876\,,
\ee
%%%%%
we can find a non-Schwarzschild static black hole.  For each such $r_0$, there
is a corresponding value $\delta=\delta^*$ of the ``non-Schwarzschild
parameter''
that yields the black-hole solution without a singularity at spatial infinity.  As $r_0$ is taken closer
and closer to the value $r_0^{\rm min}$, the required value  $\delta^*$
becomes smaller and smaller, tending to zero at $r_0=r_0^{\rm min}$.  Thus
the Schwarzschild and the non-Schwarzschild black holes ``coalesce'' as
$r_0=r_0^{\rm min}$ is approached.

  As $r_0$ is increased above $r_0^{\rm min}$, the Schwarzschild and
non-Schwarzschild black-hole solutions separate more from one another (and
in particular the required value of $\delta$ increases).
The mass of the Schwarzschild black hole is simply $\ft12 r_0$,
and thus it increases linearly as $r_0$ increases.  By constrast, the
mass of the non-Schwarzschild black hole decreases as $r_0$ increases,
until at $r_0=r_0^{\rm m=0}$ it becomes massless, where
%%%%%
\be
r_0^{\rm m=0} \approx 1.143\,.
\ee
%%%%%
(The definition of mass in higher-derivative theories was discussed in
\cite{bds,destek}.  For asymptotically-flat black holes it is just $\ft12$
the coefficient of $1/r$ in $g_{tt}$ (assuming $g_{tt}$ is normalised canonically
at infinity).)
Interestingly, if $r_0$ is increased beyond $r_0^{\rm m=0}$, one can
still obtain a non-Schwarzschild black-hole solution for an appropriate
choice of $\delta$, but now the mass is actually negative.  In other
words there is still a regular horizon, and the metric is asymptotically
flat at large distances,  but the metric function $f$ now rises above 1
as $r$ increases from $r_i$, before sinking down to 1 again in the
asymptotic region.  Figure 1 shows the masses 
of the Schwarzschild
and non-Schwarzschild black holes as a function of $r_0$, and their
masses as functions of their Hawking temperatures.  The maximum
possible mass for the non-Schwarzschild black hole, attained when
$r_0=r_0^{\rm min}$, is given by $M^{\rm max}=\ft12r_0^{\rm min} \approx
0.438$.  From the slope of $M(T)$ it can be seen that the specific heat
$C=\del M/\del T$ is negative for both black holes, and more negative
for the non-Schwarzschild black hole at a given temperature.

\begin{figure}[h!]
\begin{center}
\includegraphics[width=120pt]{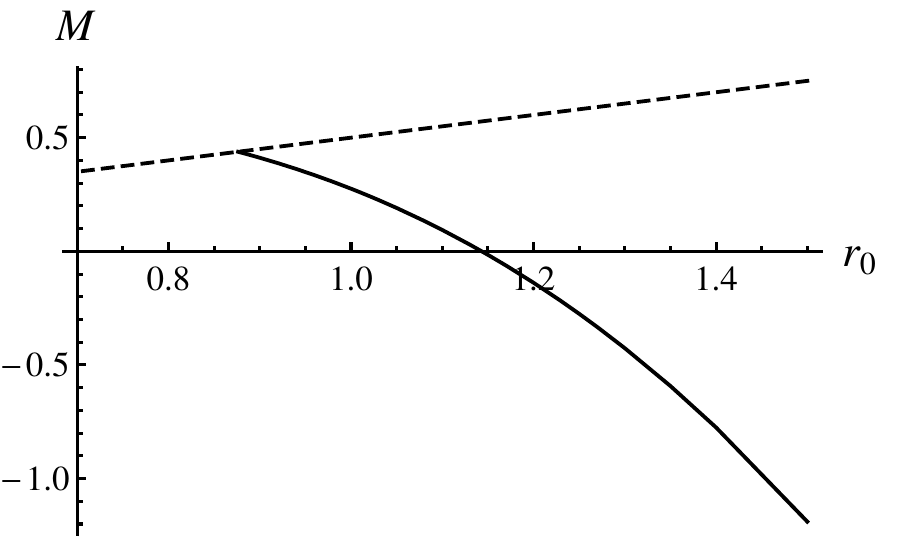}\ \ 
\includegraphics[width=120pt]{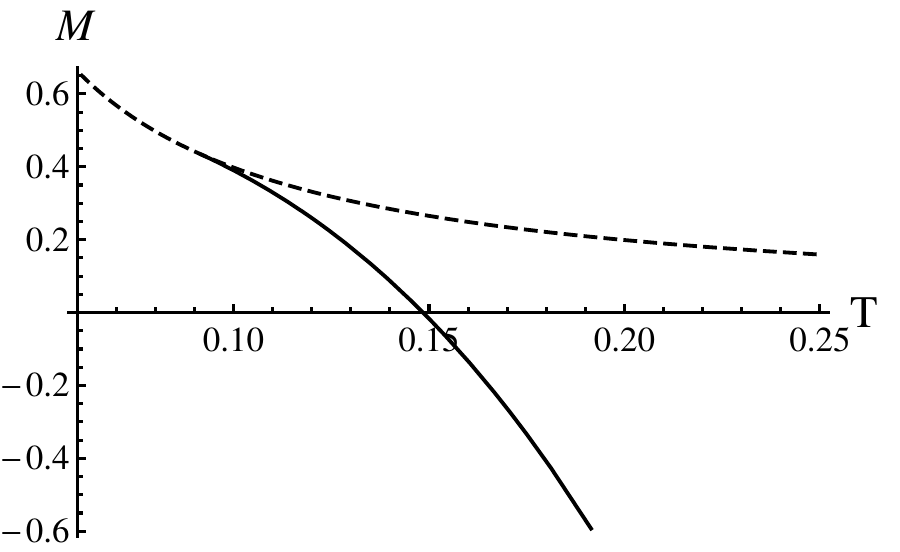}
\end{center}
\caption{{\it The masses as functions of $r_0$ (left plot),
and as functions of the Hawking temperatures (right plot)
for the Schwarzschild (dashed line) and
non-Schwarzschild (solid line) black holes.}}
\end{figure}

The plots of the metric functions $f$ and $h$ for
the examples of a positive-mass black hole with $r_0=1$, and a negative-mass
black hole, with $r_0=2$, are shown in Figure 2.
%Of course in order to obtain
%the solution with a conventionally-normalised asymptotic time coordinate,
%one would need to exploit the trivial scaling symmetry of the $h$ metric
%function, as embodied in the constant factor $c$ in (\ref{hfexp}),
%to make $h$ approach 1 at infinity.

\begin{figure}[h!]
\begin{center}
\includegraphics[width=120pt]{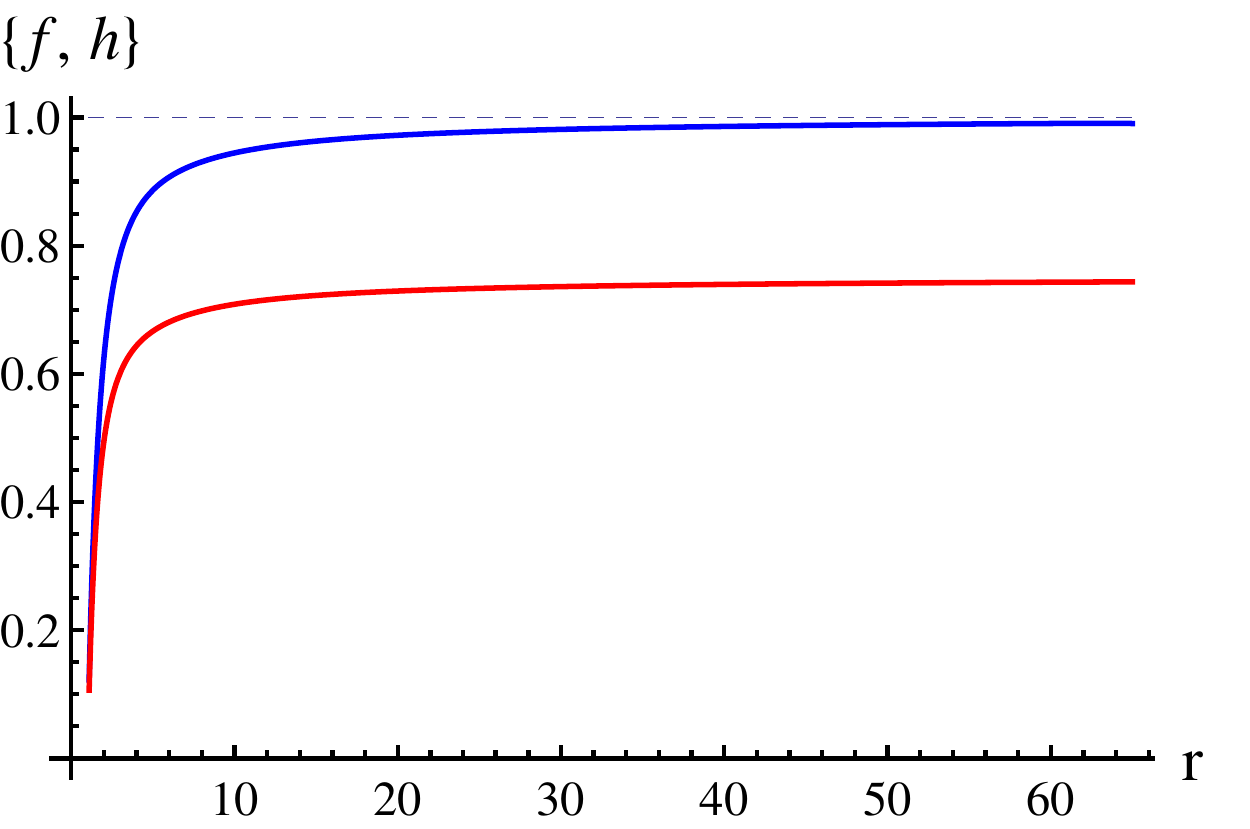}
%\end{center}
%\end{figure}
%\begin{figure}[h!]
%\begin{center}
\includegraphics[width=120pt]{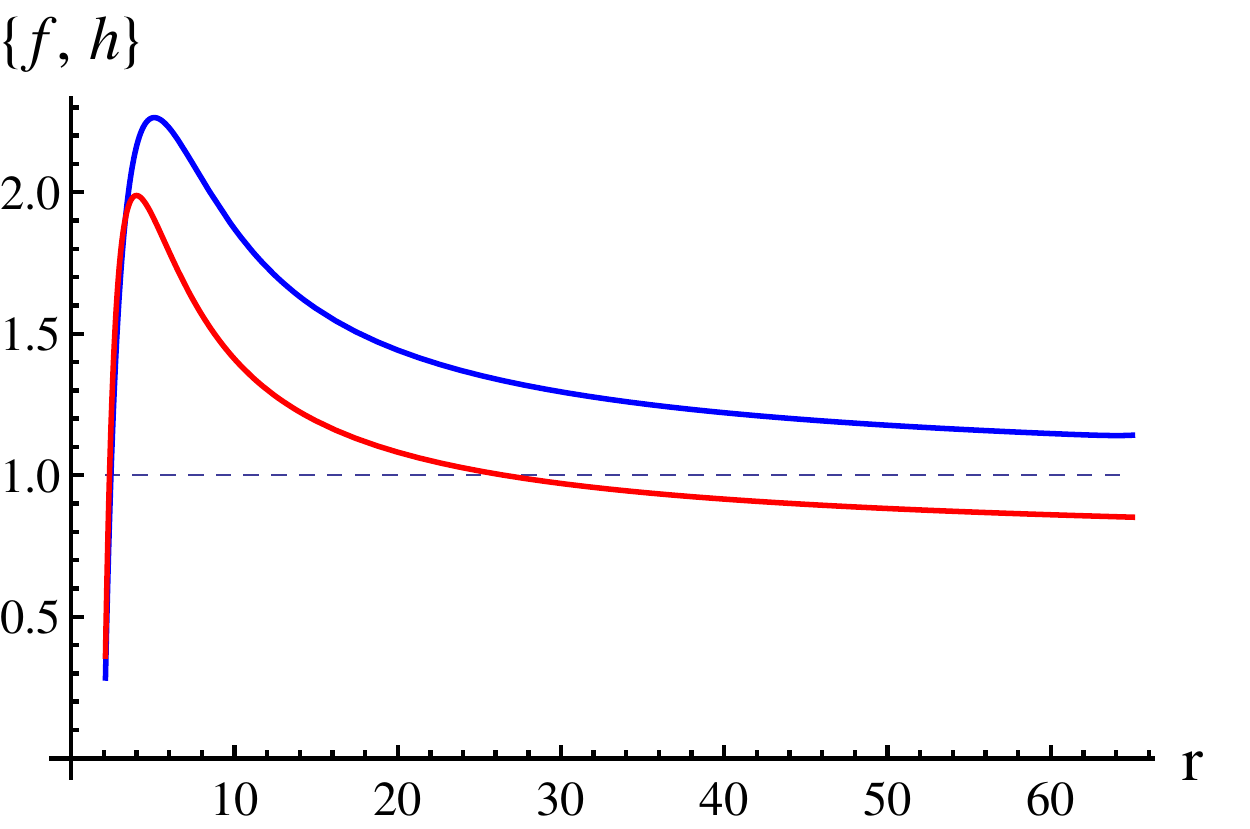}
\end{center}
\caption{{\it The non-Schwarzschild black hole for $r_0=1$ (left plot)
and $r_0=2$ (right plot). In each plot the upper curve is $f(r)$ and the
lower curve is $h(r)$.  For clarity we have chosen a rescaling of $h$ so that
it approaches $\ft34$, rather than 1, to avoid an asymptotic overlap of the
curves.}}
\end{figure}

   Having established the existence of the non-Schwarzschild black holes,
it is instructive to study some of their thermodynamic properties, and
to compare these with the properties of the Schwarzschild black holes.
In order to do this, we have collected the numerical results for a sequence of
black-hole solutions with $r_0$ in the range $r_0^{\rm min}\approx 0.876 <
r_0 < 1.5$, and then fitted the data to appropriate polynomials.  Because
we are working with a higher-derivative theory, the entropy is not
simply given by one quarter of the area of the event horizon, and instead we
need to use the formula derived by Wald \cite{wald1,iyewal}.
This has been evaluated
for the ansatz (\ref{metans}) in quadratic curvature gravities in
\cite{luentropy}, and applied to our case with $\beta=0$ and $\gamma=1$
in (\ref{HDGaction}) this gives $S=\pi r_0^2 + 4\pi\alpha (1-f_1\, r_0)=
\pi r_0^2 - 4\pi \alpha\delta^*$.  (There is a freedom to add a constant
multiple of the Gauss-Bonnet invariant to the Lagrangian, which
shifts the entropy by a parameter-independent constant
without affecting the equations of motion.  We have used this to
ensure the entropy of the Schwarzschild black hole vanishes when the
mass vanishes.)
We then find
that the mass and the temperature of these non-Schwarzschild black holes,
as a function of the entropy, take the form
%%%%%
\bea
M&\approx& 0.168 + 0.131 \, S - 0.00749 \, S^2 - 0.000139\, S^3
    +\cdots\,,\nn\\
 T &\approx& 0.131 - 0.0151 \, S - 0.000428 \, S^2 + \cdots\,.\label{MTS}
\eea
%%%%%
It can be seen that $\del M/\del S\approx 0.131 -
0.0150 \, S - 0.000417 \, S^2$, which is very close to the expression
for the temperature.  Thus the non-Schwarzschild black holes
are seen to obey the first law $dM=T dS$ to quite a high precision.  Note
that the expressions for $M$ and $T$ as a function of $S$
for the Schwarzschild black holes are very different in form, with
$M=(S/4\pi)^{1/2}$ and $T=\ft14 (\pi S)^{-1/2}$. 

\begin{figure}[h!]
\begin{center}
\includegraphics[width=120pt]{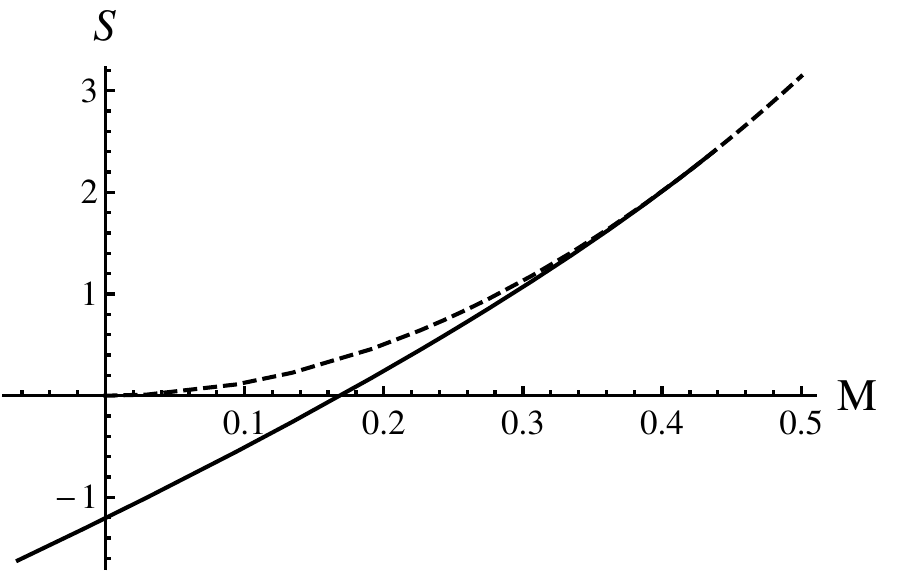}\ \  
%\end{center}
%\end{figure}
%\begin{figure}[h!]
%\begin{center}
\includegraphics[width=120pt]{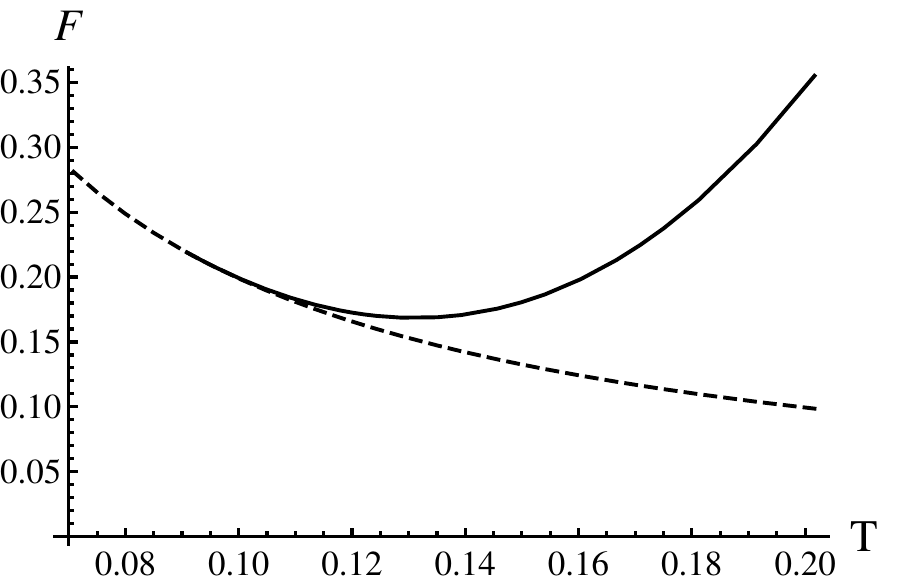}
\end{center}
\caption{{\it The first plot shows the entropy as a function of mass,
and the second shows the free energy $F=M-T S$ as a function of
$T$, for
the Schwarzschild (dashed line) and non-Schwarzschild (solid line)
black holes.}}
\end{figure}

   It is interesting to note that the entropy of the non-Schwarzschild
black hole of a given mass is always less than the entropy of the
Schwarzschild black hole of the same mass.  The two entropies
approach each other asymptotically as $r_0$ approaches $r_0^{\rm min}$.
This can be seen in the left-hand plot in Figure 3.  It is also of interest
to look at the free energy $F=M-T S$ as a function of temperature.  This
is shown in the right-hand plot in Figure 3.  It can be seen that the free
energy is always larger for the non-Schwarzschild black hole at a given
temperature, with the two curves again meeting at the lower limit when
$r_0=r_0^{\rm min}$.

   In this paper, we have used black holes to probe 
some of the consequences of interpreting the action (\ref{HDGaction}) 
as complete 
classical action in its own right.  We have seen that there exists a
second branch of static, spherically symmetric black holes, over and
above the Schwarzschild solutions. These are not Ricci flat, although they
do have vanishing Ricci scalar.  Restoring the factors of $\alpha$ and
$\gamma$ that we fixed in our numerical simulations, the second branch
of black holes have masses, which can become negative, 
bounded approximately by $M\le 0.438 \sqrt{2\alpha\gamma}$.  Thus in
a regime where $\alpha$ is small, which one might hope would correspond to
a small correction to Einstein gravity, the second branch of black holes will
be tiny, and will actually have very large curvature near the horizon, 
thus tending to
invalidate the requirement that the curvature-squared should be small.  
The fact that their mass can be negative, violating the
usual positive-mass theorem of standard Einstein gravity, indicates that 
the ghost-like nature of the 
quadratically-corrected action is becoming dominant in this regime; one 
may view the negative-mass black holes as condensates dominated by
the contribution of the ghost-like massive spin-2 modes. 
It could be viewed as a satisfactory outcome of our investigation that 
the only indications of the existence of black holes with potentially
pathological properties in the quadratic-curvature theories occur
in a regime where yet higher-order corrections, as in string theory, are 
going to be important also.  It would be interesting to obtain analytical
proofs of the existence of the numerical solutions we have found.  Although
this could be challenging, it might perhaps be easier to obtain restricted
no-hair theorems that confirmed the apparent absence of non-Schwarzschild
black holes outside the parameter range where we have found them.  

Naturally, the fourth-order equations of motion \eqref{eoms} have wider 
classes 
of solutions than the black-hole solutions
with horizons that we have considered here. These were initially 
investigated in
 \cite{stelle2} and will be given a more detailed analysis, along
with a more extensive analysis of the black hole solutions, in 
\cite{lupepost}.

\medskip
\centerline{\bf Acknowledgments}
H.L., C.N.P. and K.S.S. are grateful to the KITPC, Beijing,
for hospitality during the course of this work.  H.L. and K.S.S. also thank 
the Mitchell
Institute, and K.S.S. thanks 
the Institut des Hautes Etudes Scientifiques for hospitality.
The work of C.N.P.\
was supported in part by DOE grant  DE-FG02-13ER42020; 
the work of H.L.\
was supported in part by NSFC grants 11175269, 11475024 and 11235003; 
the work of K.S.S.\ was supported in part by the STFC under Consolidated 
Grant ST/J0003533/1 and the work of A.P.\ was supported by an STFC studentship.

\end{document}